\documentclass[twocolumn,nopacs,aps,pre,floatfix,superscriptaddress]{revtex4-2}
\pdfoutput=1 
\usepackage{amsmath,amssymb,eucal}
\usepackage{graphicx}
\graphicspath{ {images/} }
\usepackage{epstopdf}

\usepackage[colorlinks=true, urlcolor=blue, anchorcolor=blue, citecolor=blue,filecolor=blue,linkcolor=blue,menucolor=blue,pagecolor=blue]{hyperref}

\usepackage{color}
\definecolor{green}{rgb}{0,0.5,0}
\usepackage{textcomp}

\def\beq{\begin{eqnarray}}
\def\eeq{\end{eqnarray}}
\def\be{\begin{equation}}
\def\ee{\end{equation}}

\def\bm{\begin{math}}
\def\me{\end{math}}

\def\beel{\begin{eqnarray} \label}

\newcommand \bei {\begin{itemize}}
\newcommand \eei  {\end{itemize}}

\begin{document}

\title{ Aggregation in non-uniform systems with advection and localized source }

\author{R. Zagidullin}
\affiliation{Skolkovo Institute of Science and Technology, Moscow, 143026 Russia}
\affiliation{Faculty of Computational Mathematics and Cybernetics, Lomonosov Moscow State University, Moscow, 119991, Russia}
\author{A.P. Smirnov}
\affiliation{Faculty of Computational Mathematics and Cybernetics, Lomonosov Moscow State University, Moscow, 119991, Russia}
\author{S. Matveev}
\affiliation{Faculty of Computational Mathematics and Cybernetics, Lomonosov Moscow State University, Moscow, 119991, Russia}
\affiliation{Marchuk Institute of Numerical Mathematics of Russian Academy of Sciences, Moscow, 119333, Russia}
\author{N. V. Brilliantov}
\affiliation{Skolkovo Institute of Science and Technology, Moscow, 143026 Russia}
\author{P. L. Krapivsky}
\affiliation{Skolkovo Institute of Science and Technology, Moscow, 143026 Russia}
\affiliation{Department of Physics, Boston University, Boston, Massachusetts 02215, USA}
\affiliation{Santa Fe Institute, 1399 Hyde Park Road, Santa Fe, New Mexico 87501, USA}


\begin{abstract}
We explore analytically and numerically agglomeration driven by advection and localized source. The system is inhomogeneous in one dimension, viz. along the direction of advection. 
We analyze a simplified model with mass-independent advection velocity, diffusion coefficient, and reaction rates. We also examine a model with mass-dependent coefficients describing aggregation with sedimentation. For the simplified model, we obtain an exact solution for the stationary spatially dependent agglomerate densities. In the model describing aggregation with sedimentation, we report a new conservation law and develop a scaling theory for the densities. For numerical efficiency we exploit the low-rank approximation technique; this  dramatically increases the computational speed and  allows simulations of large systems. The numerical results are in excellent agreement with the predictions of our theory. 
\end{abstract}

\maketitle

\section{Introduction}
Aggregation is ubiquitous in natural systems and technology \cite{Smo16,muller1928allgemeinen,Smo17,Chandra43,krapbook,Leyvraz2003, FalkovichNature, Falkovich2006}. The size and physical nature of aggregating objects are diverse ranging from astrophysical systems, where galaxies form clusters \cite{Galaxies, oort1946gas} to everyday-life examples, such as aggregation of  aerosols in smog \cite{Friedlander,Seinfeld}, coagulation in colloidal solutions (like milk) \cite{Colloid1,Colloid2},  aggregation of red blood cells  \cite{blood}, or blood clotting \cite{bloodclott}.  Nevertheless, the agglomeration kinetics and the characteristics of these processes such as the size distribution of aggregates are rather general, they  are described by a common mathematical model. The latter is based on an infinite set of rate equations for the evolution of the cluster densities $c_k(t)$, that is, of the aggregates comprised of $k$ elementary units, the monomers. Such equations, known as Smoluchowski equations \cite{Smo16,Smo17},  read: 
\begin{eqnarray}
\label{loc:input_homo}
\frac{\partial c_k}{\partial t} = \frac{1}{2}\sum_{i+j=k}K_{i,j}c_ic_j-c_k\sum_{j\geq 1}K_{k,j}c_j. 
\end{eqnarray}
Here $K_{i,j}$ are the rate coefficients, which quantify the reaction rates of the cluster merging, $[i]+[j] \to [i+j]$. The first term in the right-hand side of Eq. \eqref{loc:input_homo} describes the increase of the concentration of clusters of size $k$ due to merging  of clusters of size $i$ and $j$ (the factor $1/2$ prevents double counting). The second term describes the decay of $c_k(t)$ due to the merging of such clusters with all other clusters or monomers.   The above equations refer to  spatially uniform systems without fluxes, sources and sinks. The densities $c_k(t)$ are the average densities, so fluctuations are ignored. 

The spatially uniform systems obeying Eqs.~\eqref{loc:input_homo} are well studied analytically and numerically for a large class of aggregation kernels, see \cite{krapbook,Leyvraz2003} for a review; for some particular aggregation kernels, exact solutions have been found. In many systems, however, the agglomeration is accompanied by fluxes entraining aggregating particles.  This refers to blood cells in vessels, milk in butter-making devices, etc.; one can also mention sedimenting atmospheric dust.  Despite the high importance of such processes for numerous applications, general studies of  spatially inhomogeneous systems with advection are still lacking. The presence of a source formally makes such systems more  complex, but occasionally it gives rise to stationary solutions that are much easier to treat. This paves a way for the  construction  of a general theory of space inhomogeneous system with aggregation. 

Hence we explore the aggregation processes with a spatially localized source of clusters of small mass. The details of the source play little role, so for concreteness, we consider the source of monomers. We address the systems which are inhomogeneous in one direction only ($x$ axis);  in all other directions ($y$ and $z$ axes) the properties of the system do not change. Then the densities $c_k(x,t)$ obey the system of partial differential equations  (PDEs)
\begin{eqnarray}
\label{loc:input}
\frac{\partial c_k}{\partial t} + V_k\,\frac{\partial c_k}{\partial x} &=& \frac{1}{2}\sum_{i+j=k}K_{i,j}c_ic_j-c_k\sum_{j\geq 1}K_{k,j}c_j \nonumber\\
&+& D_k\,\frac{\partial^2 c_k}{\partial x^2}+ J \delta_{k,1}\delta(x).
\end{eqnarray}
The first two terms on the right-hand side account, as previously,  for aggregation. The next term describes mixing due to diffusion, with the diffusion coefficient $D_k$ of a cluster of size $k$. The  last term represents the monomer source at $x=0$. The second term on the left-hand side accounts for the advection in the $x$ direction. $V_k$ states for the advection velocity of a cluster of size $k$. 

The system \eqref{loc:input} of infinitely many coupled non-linear PDEs is intractable. We are chiefly interested, however, in the large time behavior when cluster densities may become stationary. Previous results \cite{Sid89,PLK:3particle,PLK:source,Leyvraz2003,book} have been obtained when the advection  was ignored. Models with advection have been studied in \cite{Asymmetric:source,Kirone:source} in truly one-dimensional settings. In one-dimensional diffusion-controlled processes, the mean-field framework is erroneous. Here we study the physically relevant three-dimensional case where the mean-field rate equations approach is applicable  \footnote{In our models, advection  occurs along one spatial coordinate and hence the quantities vary only along that coordinate, yet the system is still three-dimensional.}. The advection  may however even simplify the analysis, as asymptotically it dominates over diffusion. These lessons are potentially applicable to other reaction processes driven by a spatially localized source \cite{PLK12a,PLK12b}.

At the same time, spatially inhomogeneous models of aggregation require the elaboration of efficient numerical methods. Even investigation of the uniform systems demands challenging calculations and leads to the progress of numerical tools \cite{osinsky2020low, boje2022stochastic}. There are relatively few numerical investigations of spatially non-uniform aggregation \cite{hackbusch2012numerical, bordas2012numerical, chaudhury2014computationally}. Such studies mostly deal with applied multicomponent systems and compare the numerical results either with experimental data \cite{bordas2012numerical, hackbusch2012numerical} or check the convergence via grid refinement \cite{chaudhury2014computationally}. Thus, there is a lack of analytical results for the case of non-uniform aggregation. In this work, we use a low-rank representation of the aggregation kernel for an acceleration of computations. This idea has been reported in \cite{matveev2015, chaudhury2014computationally, zagidullin_2017} and in our work, we follow this methodology with a slight modification of the calculation of the advection term.

The goal of the present study is to analyze the simultaneous impact of spatial inhomogeneity and advection. We focus on the stationary solutions that, as we show below, may be analyzed in the framework of spatially homogeneous Smoluchowski equations. We start from a simplified model that allows an analytical treatment and then proceed with more realistic models, for which we develop a scaling  theory. We also perform comprehensive numerical simulations. The rest of the article is organized as follows. In the next Sec. II we present analytical treatment of a simplified model with constant coefficients. In Sec. III we explore systems with mass-dependent coefficients. The analysis is performed in the context of sedimentation. Sec. IV is devoted to numerical analysis and it contains the comparison of the theoretical and simulation results. Finally, in Sec. V we summarize our findings.

\section{Mass-independent coefficients }
\label{sec:const}

Consider a toy model with mass-independent reaction rates, diffusion coefficients, and velocities:
\begin{equation}
K_{i,j}=2K, \quad D_k=D, \quad V_k=V
\end{equation}
The dimensions of the parameters $D$, $K$, $V$, and $J$ are
\begin{equation}
[D]=\frac{L^2}{T}\,, \quad [V]=\frac{L}{T}\,, \quad [K]=\frac{L^3}{T}\,, \quad [J]=\frac{1}{L^2 T}
\end{equation}
where $L$ and $T$ denote the units of length and time.
In writing the dimensions of $K$ and $J$ we have assumed that the process occurs in  three dimensions; the general expressions are $[K]=L^d/T$ and $[J]=1/(L^{d-1} T)$. In the following we use $K/D$ as the unit of length and $K^2/D^3$ as the unit of time. Using these units is equivalent to setting $K=1$ and $D=1$. The governing equations read
\begin{equation}
\label{VJ-ck}
V c_k' = \sum_{i+j=k}c_ic_j-2c_k c + c_k''+J \delta_{k,1}\delta(x)
\end{equation}
in the stationary regime. Here  $c=\sum_{k\geq 1}c_k$ is the total cluster density and prime denotes the derivative with respect to $x$. The velocity $V$ and the source strength $J$ in \eqref{VJ-ck} are dimensionless quantities. To avoid cluttering formulas we denoted them by the same letters, although they are given by $VK/D^2$ and  $JK^4/D^5$ in terms of the original quantities. The dimensional quantity ${\rm Pe}= VK/D^2$ is particularly important, it is the P\'{e}clet number. Indeed, according to the of definition,  of ${\rm Pe}$,  this  is the ratio of the product of a  characteristic velocity $V$ and characteristic length $K/D$ over the characteristic diffusion coefficient $D$, which yields the above quantity. Note  that in the adopted  time and length units the dimensionless velocity coincides with the P\'{e}clet number, that is, ${\rm Pe}=V$. The P\'{e}clet number quantifies the relative importance of advection as compared to the diffusion. For large ${\rm Pe}$ the diffusion is negligible. For the system with constant coefficients the P\'{e}clet number is a quantity, which characterizes all clusters. 

The total cluster density obeys
\begin{equation}
\label{VJ-c}
V c' = c''-c^2+J\delta(x)
\end{equation}
Before analyzing this equation we consider the case without advection  ($V=0$).

\subsection{No advection }

Since due to symmetry, $c(x)=c(-x)$, it suffices to solve
\begin{equation}
\label{J-c}
c''-c^2 = 0
\end{equation}
for $x>0$, with the source providing the boundary condition
\begin{equation}
\label{BC}
2c'(+0)=-J
\end{equation}
obtained by integrating near the origin. The solution of \eqref{J-c}--\eqref{BC} reads \cite{Sid89}
\begin{equation}
c(x) = \frac{6}{(|x|+\ell)^2}\,, \quad \ell = \left(\frac{24}{J}\right)^{1/3}.
\label{c:noadv}
\end{equation}
In particular
\begin{equation}
\label{c0}
c(0) =\frac{3}{2}\,  \left(\frac{J}{3}\right)^{2/3}.
\end{equation}

To determine stationary cluster densities one can employ the generating function technique. It is convenient to use the generating function
\begin{equation}
\label{GF:def}
\mathcal{C}(x,z) =\sum_{k\geq 1} \big(1-z^k\big)\,c_k(x)
\end{equation}
This generating function satisfies an equation mathematically identical to Eq.~\eqref{J-c}, viz. $\mathcal{C}'' - \mathcal{C}^2=0$. The boundary condition is almost the same as before
\begin{equation}
\label{BC-z}
2\mathcal{C}'(+0,z)=-J(1-z)
\end{equation}
Therefore the solution reads \cite{Sid89,PLK:source,Leyvraz2003}
\begin{equation}
\label{GF:sol}
\mathcal{C}(x,z) = \frac{6}{(|x|+\ell(1-z)^{-1/3})^2}
\end{equation}
The densities remain finite at the origin. Specializing \eqref{GF:sol} to $x=0$ and expanding in powers of $z$ one finds
\begin{equation}
\label{ck:origin}
c_k(0) = \left(\frac{J}{3}\right)^{2/3} \frac{\Gamma\big(k-\frac{2}{3}\big)}{\Gamma\big(\frac{1}{3}\big) \Gamma(k+1)},
\end{equation}
where $\Gamma(x)$ is the Gamma-function. 

We also display a few exact cluster densities
\begin{equation*}
\begin{split}
c_1(x) & = \frac{4\ell}{(|x|+\ell)^3}\\
c_2(x) & = \frac{8}{3}\, \frac{\ell}{(|x|+\ell)^3} - \frac{2\ell^2}{(|x|+\ell)^4}\\
c_3(x) & =  \frac{56}{27}\, \frac{\ell}{(|x|+\ell)^3} - \frac{8}{3}\, \frac{\ell^2}{(|x|+\ell)^4}
+ \frac{8}{9}\,\frac{\ell^3}{(|x|+\ell)^5}
\end{split}
\end{equation*}
and the asymptotic
\begin{equation}
c_k(x)\simeq \frac{12\ell}{\Gamma\big(\frac{1}{3}\big)}\,k^{-2/3} |x|^{-3}
\label{ck:noadv}
\end{equation}
valid when $1\ll k\ll (|x|/\ell)^3$. These results follow from the exact expression \eqref{GF:sol} for the generating function.

Total cluster numbers
\begin{equation}
C_k = \int_{-\infty}^\infty dx\, c_k(x), \qquad C =  \int_{-\infty}^\infty dx\, c(x)
\end{equation}
also become stationary in the large time limit. Using \eqref{cx} we find the total number of clusters
\begin{equation}
\label{Cx}
C(x) = \frac{12}{\ell}.
\end{equation}
Using \eqref{GF:sol} we determine the generating function
\begin{equation}
\label{Cz:def}
\mathcal{C}(z) =\sum_{k\geq 1} \big(1-z^k\big)\,C_k = \frac{12}{\ell}\,(1-z)^{1/3}
\end{equation}
from which
\begin{equation}
\label{Ck}
C_k = \frac{4}{\ell}\,  \frac{\Gamma\big(k-\frac{1}{3}\big)}{\Gamma\big(\frac{2}{3}\big) \Gamma(k+1)}.
\end{equation}

Not all quantities become stationary in the long time limit. The total mass
\begin{equation}
\label{mass}
M(t) = Jt
\end{equation}
is the most obvious example. The mass density
\begin{equation}
m(x,t) = \sum_{k\geq 1}k c_k(x,t)
\end{equation}
is also non-stationary. It satisfies
\begin{equation}
\label{mass-eq}
\frac{\partial m}{\partial t} = \frac{\partial^2 m}{\partial x^2}+ J\,\delta(x)
\end{equation}
which is solved to yield
\begin{equation}
\label{mass-sol}
m(x,t)= J\sqrt{t}\left[\pi^{-1/2}\,e^{-\xi^2}-|\xi|\,\text{Erfc}(|\xi|)\right]
\end{equation}
where $\xi=x/\sqrt{4t}$. In particular, the mass density at the origin grows as
\begin{equation}
\label{mass-origin}
m(0,t)= J\sqrt{\frac{t}{\pi}}.
\end{equation}
Thus the densities at the origin are stationary and described by Eq.~\eqref{ck:origin} up to a certain crossover mass $K(t)$, while for $k>K(t)$ the densities quickly vanish. Using \eqref{ck:origin} and the asymptotic $\frac{\Gamma\big(k-\frac{2}{3}\big)}{\Gamma(k+1)}\simeq k^{-5/3}$ we obtain
\begin{equation*}
m(0,t) \sim \sum_{k=1}^{K} k c_k(0) \sim J^{2/3}\sum_{k=1}^{K} k^{-2/3}\sim  J^{2/3} K^{1/3}
\end{equation*}
consistent with \eqref{mass-origin} when the crossover mass scales as
\begin{equation}
K \sim J t^{3/2}
\end{equation}

\subsection{The impact of advection }

The case of a practical importance (e.g. sedimentation of aggregating particles)  corresponds to $V>0$ in the half-space $x\geq 0$, when clusters are driven away (say by gravity) from the source.  The proper extension to the entire space is to assume that $V<0$ when $x<0$. The solution is symmetric, $c(x)=c(-x)$, so it suffices to consider the $x\geq 0$ half-space. 

Thus we must solve
\begin{equation}
\label{c:eq}
V c' = c''-c^2
\end{equation}
for $x>0$ subject to the boundary condition \eqref{BC}.  Equation \eqref{c:eq} does not admit an analytical solution, so we write the stationary cluster density $c(x)$ in a formal form
\begin{equation}
\label{c:JV}
c(x) = \Phi(x; J, V)
\end{equation}
emphasizing the dependence on the parameters. Specializing \eqref{c:JV} to $x=0$ one gets
\begin{equation}
\label{c0:JV}
c(0) = F(J, V), \qquad F(J, V)=\Phi(0; J, V)
\end{equation}
We know $\Phi(x; J, V)$ and $F(J, V)$ when $V=0$, 
\begin{equation*}
\label{FJ}
 \Phi(x;J,0)= \frac{6}{(|x|+ \ell)^2}, \qquad
  F(J, 0) = \frac{6}{\ell^2} = \frac{3}{2}\,  \left(\frac{J}{3}\right)^{2/3}. 
\end{equation*}

When $x\gg V^{-1}$, the advection dominates over diffusion. Thus \eqref{c:eq} simplifies to $V c' \simeq -c^2$ from which
\begin{equation}
\label{c-asymp}
c(x) \simeq V\,x^{-1}\qquad\text{for}\quad x\gg V^{-1}
\end{equation}
Restoring dimensional units, one re-writes \eqref{c-asymp} as
\begin{equation}
\label{c-VK}
c(x) \simeq \frac{V}{K}\,x^{-1} \qquad\text{for}\quad x\gg \frac{D}{V}
\end{equation}
Remarkably, this behavior is {\em independent} on the strength $J$ of the source which drives the system: As long as $J>0$, the leading asymptotic behavior \eqref{c-asymp} is the same. We also emphasize that the $x^{-1}$ decay is much slower than the $x^{-2}$, see \eqref{cx}, arising in the no-advection case.

The density of monomers satisfies (we return to dimensionless variables)
\begin{equation}
\label{c1:eq}
V c_1' = c_1''-2c c_1
\end{equation}
and the boundary condition
\begin{equation}
\label{BC:1}
2c_1'(+0)=-J
\end{equation}
The asymptotic behavior is found from $V c_1' \simeq -2c c_1$ which simplifies [cf. Eq.~\eqref{c-asymp}] to $c_1' \simeq -2c_1/x$ leading to
\begin{equation}
c_1(x) \simeq \frac{A_1}{x^2}\quad\text{for}\quad x\gg V^{-1}
\end{equation}
Similarly $c_k \simeq A_k/x^2$ for $x\gg V^{-1}$. To determine $A_k$ we first simplify the governing equations \eqref{VJ-ck} to
\begin{equation}
\label{V-ck}
V c_k' = \sum_{i+j=k}c_ic_j-2c_k c
\end{equation}
valid when $x\gg V^{-1}$. These are Smoluchowski equations for pure aggregation with mass-independent merging rates \cite{Smo17,Chandra43,Drake}.
In the scaling limit $k\to\infty$ and $x\to\infty$ with ratio $k/x$ kept finite, the solution has the well-known scaling form \cite{Leyvraz2003,book}
\begin{equation}
\label{ck-scal}
c_k(x) =  \frac{A}{x^2}\,e^{-kB/x}
\end{equation}
To determine $A$ and $B$ we use the sum rules
\begin{subequations}
\begin{align}
\label{c-V}
c(x)&=\sum_{k\geq 1}c_k\simeq \frac{V}{x}\\
\label{m-JV}
m(x)&=\sum_{k\geq 1}kc_k \simeq \frac{J}{2V}
\end{align}
\end{subequations}
valid when $x\gg V^{-1}$. The sum rule \eqref{c-V} is the re-statement of Eq.~\eqref{c-asymp}, the sum rule \eqref{m-JV} is established below. Using \eqref{ck-scal} we recover the sum rules if $A=2V^3/J$ and $B=2V^2/J$. Thus
\begin{equation}
\label{ck-scaling}
c_k(x) =  \tfrac{2V^3}{J}\,x^{-2}\exp\!\big[-\tfrac{2V^2}{J}\,\tfrac{k}{x}\big]
\end{equation}

To derive the sum rule \eqref{m-JV} we rely on equation
\begin{equation}
\label{mass-eq-V}
\frac{\partial m}{\partial t} + V\frac{\partial m}{\partial x}= \frac{\partial^2 m}{\partial x^2}+ J\,\delta(x)
\end{equation}
for the mass density. In the interesting range $x\gg V^{-1}$ we can again neglect the diffusion:
\begin{equation}
\label{mass-V}
\frac{\partial m}{\partial t} + V\frac{\partial m}{\partial x}=0
\end{equation}

The solution is a constant in the interval $(-Vt, Vt)$ apart from the central region $|x|<V^{-1}$ where diffusion matters;
the mass density quickly vanishes outside the $(-Vt, Vt)$ interval. The total mass is $2m Vt$ up to $O\big(\sqrt{t}\big)$ corrections due to the boundary layers near the boundaries of the $(-Vt, Vt)$ interval where the mass density quickly vanishes. Comparing with the exact value \eqref{mass} of the total mass we get $m=\frac{J}{2V}$ as we stated in \eqref{m-JV}.

Finally, we re-write our basic findings in terms of the original dimensional variables. The sum rule \eqref{c-V} becomes \eqref{c-VK}. The sum rule \eqref{m-JV} maintains its form up to the crossover length:
\begin{equation}
m(x)\ \simeq \frac{J}{2V}\qquad\text{for}\quad x\gg \frac{D}{V}
\end{equation}
The scaling expression \eqref{ck-scaling} for the cluster-mass distribution becomes
\begin{equation}
\label{ck-JKV}
c_k(x) =  \tfrac{2V^3}{J K^2}\,x^{-2}\exp\!\big[-\tfrac{2V^2}{J K}\,\tfrac{k}{x}\big] \qquad\text{for}\quad x\gg \frac{D}{V}
\end{equation}

The total number of clusters
\begin{eqnarray*}
C_k &\simeq& \frac{4V^3}{J K^2}\int_{D/V}^\infty \frac{dx}{x^2}\,\exp\!\big[-\tfrac{2V^2}{J K}\,\tfrac{k}{x}\big]\\
&=& \frac{2V}{K}\,k^{-1}\left[1-e^{-\frac{2V^3}{J KD}k}\right]
\end{eqnarray*}
When $k\gg JKD/V^3$, the asymptotic simplifies to
\begin{equation}
C_k \simeq \frac{2V}{K}\,k^{-1}
\end{equation}

\section{Mass-dependent coefficients and sedimentation}

Consider the system \eqref{loc:input} with mass-dependent coefficients. Here we address an important case of particle sedimentation under gravity. Suppose that the monomers are released with a constant rate $J$ on the gas-liquid (air-water) interface. They go down subjected to gravity and merge forming clusters. The clusters move with the velocity growing with their  mass; clusters also diffuse. We consider dilute systems where aggregation events are binary and the motion of each cluster is essentially independent of other clusters apart from merging events. A faithful description of the merging events is a challenging problem beyond the scope of our work. We merely assume that this process is sufficiently fast and newly formed clusters quickly reach their final velocity; the terms ``fast'' and ``quickly'' mean that the characteristic times of these processes are much smaller than the typical time between collisions for a given cluster. The Reynolds number associated with the motion of every cluster is assumed to be small assuring that the system is in the Stokes regime and particles move with steady velocities.

If $m_1$ and $r_1$ denote the mass and the radius of the monomer, the cluster of size $k$ has mass $m_1k$ and radius $R_k=r_1k^{1/3}$. If the density of monomers is $\rho$ while the density of liquid is $\rho_l$, the force acting on a particle due to the buoyancy and gravity is $(4\pi/3) k\, r_1^3(\rho-\rho_l)g$. Moving in fluid the particle also experiences viscous force.  In the Stokes regime it is proportional to the velocity $V_k$ and the the friction coefficient $\gamma_k = 6 \pi \eta  R_k = 6\pi \eta r_1 k^{1/3}$,   where $\eta$ is the fluid viscosity. That is, this force reads, $-\gamma_k V_k$ \cite{OTB89,krapbook}. A  steady  velocity corresponds to the vanishing total force: $(4\pi/3) k\, r_0^3(\rho-\rho_l)g =6\pi \eta r_0 k^{1/3} V_k$. Thus
\begin{equation}
\label{Vk}
V_k=V_1 k^{2/3};  \quad \qquad V_1 = \frac29 r_1^2 g (\rho -\rho_l).
\end{equation}

The diffusion coefficient for a cluster of size $k$ obeys the Stokes-Einstein relation,  $D_k=k_BT/\gamma_k$, where $k_B$ is the Boltzmann constant and $T$ is temperature \cite{OTB89,krapbook}. This results in the following size-dependence of $D_k$,

\begin{equation}
\label{Dk}
 D_k= D_1 k^{-1/3} ; \quad \qquad D_1= \frac{k_BT}{ 6 \pi \eta r_1} .
\end{equation}
Merging occurs mostly due to the ballistic motion, when one particle of size $R_i$ falling with the velocity $V_i$ collides with another particle  of size $R_j$, falling with a smaller velocity $V_j$. The collision  cross-section and relative velocity are respectively $\pi (R_i+R_j)^2 =\pi r_1^2 (i^{1/3} +j^{1/3})^2$ and  $|V_i - V_j| = V_1 |i^{2/3} -j^{2/3}|$. Hence the ballistic rate kernel reads: 
\begin{equation}
\label{Kball}
K_{ij}^{\rm ball} = K_1^{\rm ball}\left( i^{1/3} +j^{1/3} \right)^2 |i^{2/3}-j^{2/3}|, 
\end{equation}
where $K_1^{\rm ball} = \pi r_1^2V_1$; as before, we put $K_1^{\rm ball}=1$. Noteworthy,  the ballistic rates \eqref{Kball} are similar to the Smoluchowski aggregation rates  for shear flow  \cite{JCEJ}. 

The diffusion-controlled aggregation rates are  given by the Brownian kernel, $K_{ij}^{\rm Br}= 4\pi (D_i+D_j)(R_i+R_j)$, that is,
\begin{equation}
\label{Brown:kernel}
K^\text{Br}_{i,j} = K^{\rm Br}\, \big(i^{1/3}+j^{1/3}\big)\big(i^{-1/3}+j^{-1/3}\big), 
\end{equation}
with $K^{\rm Br }= 2k_BT/(3\eta)$.
Both these kernels are homogeneous, $K_{ai,aj}=a^\lambda K_{i,j}$, with $\lambda=4/3$ for the ballistic kernel \eqref{Kball} and $\lambda=0$ for the Brownian kernel \eqref{Brown:kernel}. Thus, the ballistic kernel rapidly increases with masses of the aggregates, and this provides an extra justification for using it and ignoring the merging due to diffusion  \footnote{For $i=j$ however $K_{ij}^{\rm ball} =0$, that is, the diffusion mechanism may not be neglected. Hence, for $i=j$ we use $K^\text{Br}_{i,i}$ from Eq. \eqref{Brown:kernel}, assuming that $K_1^{\rm Br }= 1$.} 

Suppose the densities become stationary in the long-time limit. In this situation
\begin{equation}
\label{Smol-sed}
\frac{dc_k}{dx} = \frac{1}{2}\sum_{i+j=k}\frac{K^\text{ball}_{i,j}}{(i+j)^{2/3}}\,c_ic_j-c_k \sum_{j\geq 1} \frac{K^\text{ball}_{k,j}}{k^{2/3}}\,c_j .
\end{equation}
These equations may be written in a form  similar to the Smoluchowski equations
\begin{equation}
\label{Smol-K}
\frac{dc_k}{dx} = \frac{1}{2}\sum_{i+j=k}K_{i,j}^{(1)}c_ic_j-c_k \sum_{j\geq 1}K_{k,j}^{(2)}c_j
\end{equation}
with the kernels
\begin{equation}
\label{KK}
K_{i,j}^{(1)} = \frac{K^\text{ball}_{i,j}}{(i+j)^{2/3}}; \qquad K_{i,j}^{(2)} = \frac{K^\text{ball}_{i,j}}{i^{2/3}}.
\end{equation}
Both kernels $K_{i,j}^{(1/2)}$ are homogeneous with the homogeneity index $\lambda =2/3<1$. The second homogeneity index $\nu$ defined via
\begin{equation}
K_{1,j}^{(1)} \sim j^{\nu_1}; \qquad \quad 
K_{j,1}^{(2)}\sim j^{\nu_2} \quad\text{when}\quad j\to\infty.
\end{equation}
Here $\nu_1=\nu_2=\nu=2/3$ for the both kernels $K_{i,j}^{(1)}$ and $K_{i,j}^{(2)}$. Therefore, if a stationary solution exists (it is not an easy task to prove this), the gelation does not occur \cite{dom,hez,spouge,van87,Colm:IG,Leyvraz2003}. Here the spatial analogue of gelation is meant, where all mass is accumulated either at $x=0$ (the analogue of instantaneous gelation \cite{BK91}) or at   $x=x_0 < \infty$.  This allows to develop a scaling approach, in analogy with the conventional Smoluchowski equations. 

Mass conservation is a chief property of Smoluchowski equations, which underpins the scaling. In our setting with spatial coordinate $x$ instead of time, the mass conservation would imply that mass density is spatially uniform. This is not true in our case. Below we derive $m(x) \sim x^{-2/3}$  which is clearly not spatially uniform. Fortunately, there exists a different conservation law, that is, the $5/3$-moment $M_{5/3}=\sum_{k\geq 1}k^{5/3} c_k(x)$ is spatially uniform. Indeed, multiplying
\begin{equation}
\label{Smol-sediment}
k^{2/3} \frac{dc_k}{dx}  = \frac{1}{2}\sum_{i+j=k}K^\text{ball}_{i,j}\,c_ic_j-c_k \sum_{j\geq 1} K^\text{ball}_{k,j}\,c_j
\end{equation}
by $k$ and summing over all $k\geq 1$ we obtain
\begin{eqnarray*}
\frac{dM_{5/3}}{dx}=\frac{1}{2}\sum_{i\geq 1}\sum_{j\geq 1}(i+j)K^\text{ball}_{i,j}c_ic_j-\sum_{k\geq 1}\sum_{j\geq 1}kK^\text{ball}_{k,j}c_jc_k .
\end{eqnarray*}
The right-hand side vanishes for any symmetric  kernel, $K_{i,j}=K_{j,i}$, so $M_{5/3}=\text{const}$. Note that this conservation law is valid for $x\gg 1/V$ (as we neglect the diffusion term) and for $x\ll X(t)$, where $X(t)$ is defined below; also $t$ should be large enough, so that the system has become stationary in this interval of $x$. 

For Smoluchowski equations, the scaling solution has the form $c_k(x) = s^{-2}\Phi(k/s)$, so the mass density is uniform:
\begin{equation*}
\sum_{k\geq 1} kc_k(x) = \int_0^\infty d\xi\, \xi\,\Phi(\xi)=\text{const}
\end{equation*}
To ensure the spatial uniformity of $M_{5/3}$ we now choose scaling solution in the form $c_k(x) = s^{-(1+5/3)}\Phi(k/s)$:
\begin{equation*}
M_{5/3}=\sum_{k\geq 1} k^{5/3}c_k(x) = \int_0^\infty d\xi\, \xi^{5/3}\,\Phi(\xi)=\text{const}.
\end{equation*}
We also make the standard assumption (justified below) that the average cluster size scales algebraically with $x$, so the scaling form is
\begin{equation}
\label{scaling-sed}
c_k(x) = x^{-8b/3}\Phi(\xi), \qquad \xi=\frac{k}{x^b} .
\end{equation}

The exponent $b$ is fixed by consistency of the scaling ansatz with the averred equations \eqref{Smol-sediment}. Plugging \eqref{scaling-sed} into \eqref{Smol-sediment} we find that the left-hand side scales as $x^{-2b-1}$, while both gain and loss terms in the right-hand side scale as  $x^{-3b}$. Hence $b=1$ and the  scaling form \eqref{scaling-sed} becomes
\begin{equation}
\label{scaling-sediment}
c_k(x) = x^{-8/3}\Phi(\xi), \qquad \xi=\frac{k}{x}
\end{equation}
Plugging \eqref{scaling-sediment} into formulae for the cluster and mass densities, $c(x)=\sum_{k\geq 1}c_k(x)$ and $m(x)=\sum_{k\geq 1}k c_k(x)$, and replacing summation by integration we arrive at 
\begin{subequations}
\begin{align}
\label{cx}
&c(x) = C_0 x^{-5/3}\,, \qquad C_0 =  \int_0^\infty d\xi\, \Phi(\xi)\\
\label{mx}
&m(x) = C_1 x^{-2/3}\,, \qquad C_1 =  \int_0^\infty d\xi\, \xi \Phi(\xi)
\end{align}
\end{subequations}
We do not know the amplitudes $C_0, C_1$ as the scaled mass density is not known analytically.

Let us probe the asymptotic behavior of the monomer density. Specifying \eqref{Smol-sediment} to $k=1$ we obtain
\begin{equation}
\label{Smol-sed-mon}
\frac{d c_1}{dx} =-c_1 \sum_{j\geq 1} \,c_j  \big(j^{1/3}+1\big)^2\, \big(j^{2/3}-1\big)
\end{equation}
Plugging the scaling form \eqref{scaling-sediment} into \eqref{Smol-sed-mon}  and replacing summation by integration gives
\begin{equation*}
\frac{d c_1}{dx} =-\tfrac{2}{3}Ax^{-1/3} c_1, \quad A= \frac{3}{2}\int_0^\infty d\xi\,\xi^{4/3}\Phi(\xi)
\end{equation*}
from which
\begin{equation}
\label{mon-sed}
c_1\sim e^{-Ax^{2/3}}
\end{equation}

Let us assume that the scaling form \eqref{scaling-sed} holds for monomers. The corresponding value of the scaling variable is
$\xi=x^{-1}$ and \eqref{scaling-sediment} is compatible with \eqref{mon-sed} if
\begin{equation}
\label{Phi-sed}
\Phi\sim e^{-A\xi^{-2/3}} \qquad\text{as}\quad \xi\to 0
\end{equation}
Plugging the scaling form \eqref{scaling-sediment} into \eqref{Smol-sed} we obtain the governing equation for the scaled mass distribution:
\begin{equation}
\label{scaled-mass}
-[\tfrac{8}{3}\Phi+\xi\Phi']\xi^{2/3}=G-L
\end{equation}
with
\begin{eqnarray}
\label{GL}
G &=& \frac12 \int_0^{\xi}d \eta K(\eta,\xi-\eta) \Phi(\eta) \Phi(\xi-\eta) \nonumber \\
L&=& \Phi(\xi)  \int_0^{\infty }d \eta K(\eta,\xi) \Phi(\eta),  
\end{eqnarray}
and where we take into account that the kernel 
\begin{equation}
    \label{Kxinu}
    K(\eta,\xi)= \left(\eta^{1/3} + \xi^{1/3} \right)^2 \left| \eta^{2/3} - \xi^{2/3} \right| 
\end{equation}
has the homogeneity exponent $\lambda =4/3$, that is, $K(x\eta,x\xi)= x^{4/3}K(\eta,\xi)$. 

In the $\xi\to\infty$ limit, the leading behavior of the scaled mass distribution is expected to be exponential up to an algebraic pre-factor:
\begin{equation}
\label{Phiexp}
\Phi \sim \xi^{-\theta} e^{-B\xi}
\end{equation}
This choice has been advocated in the realm of scaling theory \cite{van88}  and it also applies to our situation. Inserting \eqref{Phiexp} into \eqref{scaled-mass}-\eqref{GL} we find that the exponential factors are the same, so we must compare algebraic factors in the $\xi \to \infty$ limit. The left-hand side of \eqref{scaled-mass} scales as $\xi^{5/3-\theta}$, while $G\sim \xi^{7/3-2\theta}$ and $L\sim \xi^{4/3-\theta}$. Thus $L$ is negligible compared to the left-hand side. Equating the left-hand side and $G$ gives  $\theta=2/3$. Summarizing, the extremal behaviors of the scaled mass distribution are
\begin{equation}
\label{Philast}
\Phi \sim
\begin{cases}
\xi^{-2/3}\, e^{-B\xi}  & \xi\gg 1\\
e^{-A\xi^{-2/3}}           & \xi\ll 1
\end{cases}
\end{equation}
At any finite time $t$, the above description is valid up to a certain length $X(t)$. Using an estimate
\begin{equation}
M\sim \int_0^X dx\,m(x) \sim X^{1/3}
\end{equation}
and $M(t)=Jt$ we estimate the crossover length
\begin{equation}
X(t) \sim (J t)^{3}
\end{equation}
Note that the conservation of the moment $M_{5/3}$ for the interval $V^{-1} \ll x \ll X(t)$ and the associated scaling \eqref{scaling-sediment}  hold for any symmetric kernel in Eq. \eqref{Smol-sediment} (not only for the ballistic one, $K_{i,j}^{\rm ball}$). Hence all the above results are valid for any symmetric kernel $K_{i,j}$ with the same homogeneity exponent $\lambda=4/3$

\section{Numerical simulations }
\label{sec:numer}

\begin{figure}
\begin{center}
\includegraphics[width=8cm, height=8.0cm]{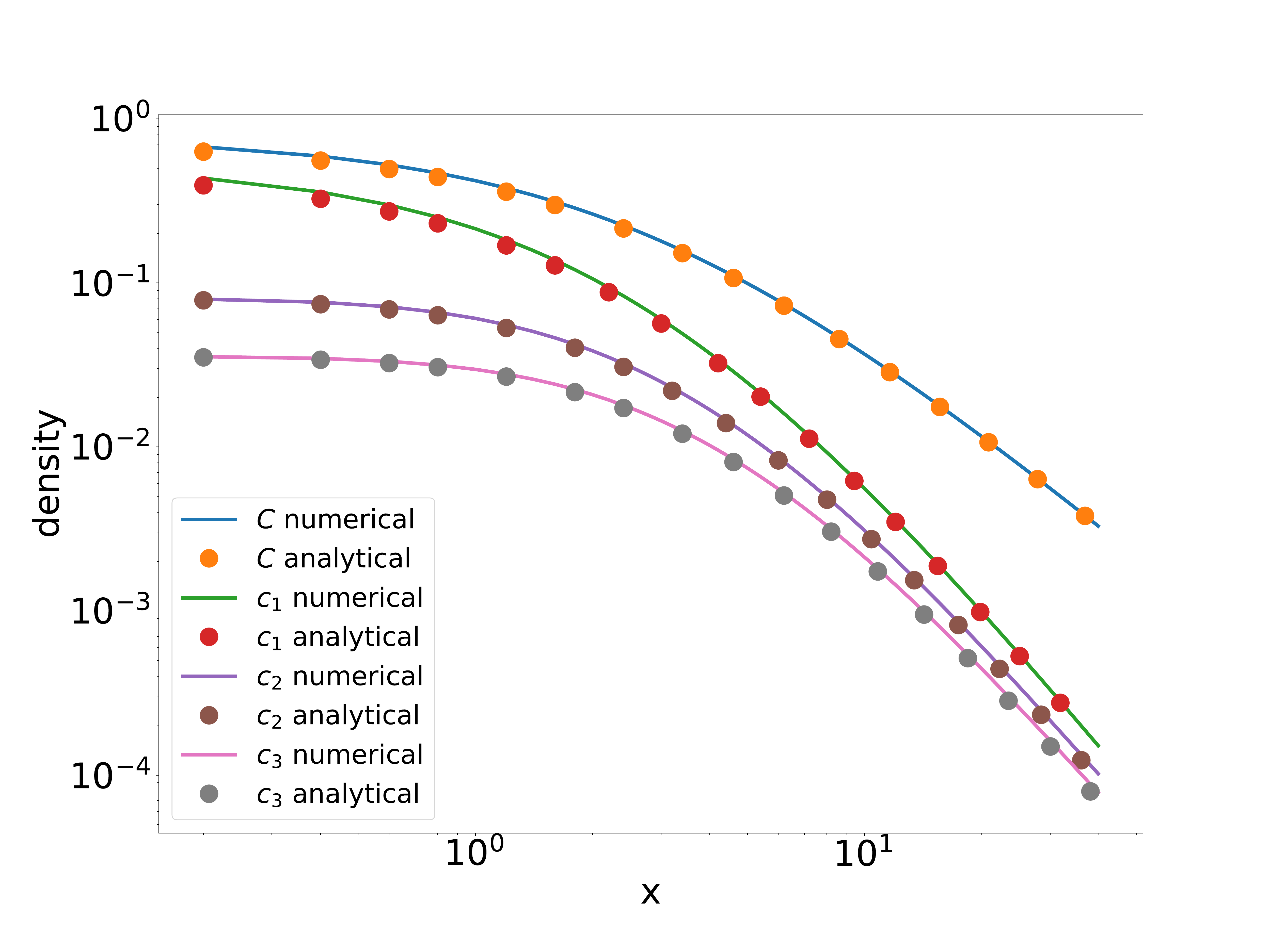}
\end{center}
\caption{Steady-state solutions ($t = 200$) to diffusion-coagulation system \eqref{loc:input} with constant coefficients. The analytical results for $c_k(x)$, Eq. \eqref{ck:noadv}, and $c(x)$, Eq. \eqref{c:noadv} (dots) are compared with the numerical solution of Eq. \eqref{loc:input} (lines) for $D=1$, $K=1$, $J=1$ and $V=0$.}
\label{diffusion}
\end{figure}

\begin{figure}
\begin{center}
\includegraphics[width=8cm, height=8.0cm]{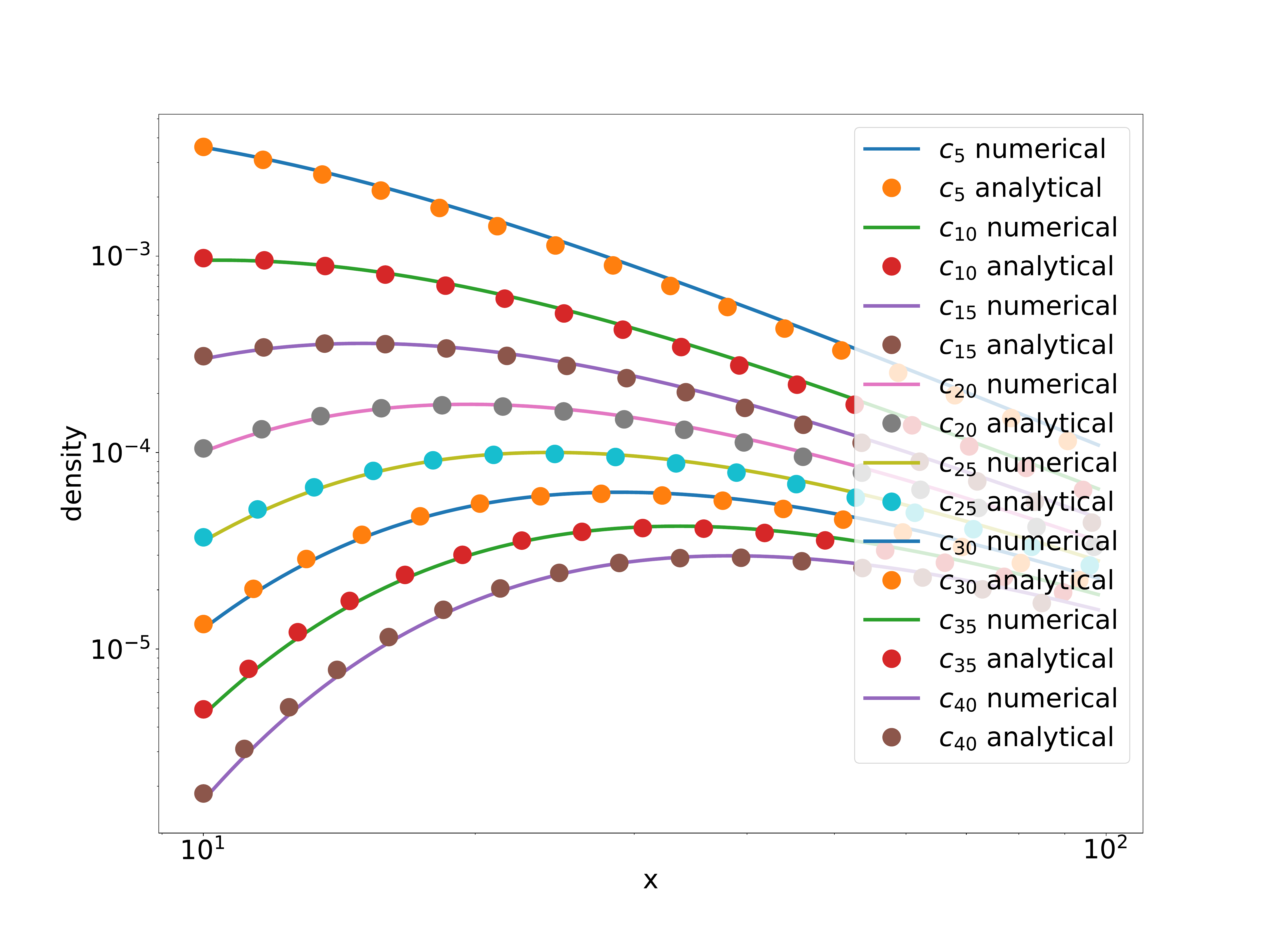}
\end{center}
\caption{Steady-state solutions ($t = 200$) to diffusion-advection-coagulation system \eqref{loc:input} with constant coefficients. The analytical results for $c_k(x)$, Eq. \eqref{ck-scaling}, and $c(x)$, Eq. \eqref{c-VK} (dots) are compared with the numerical solution of Eq. \eqref{loc:input} (lines) for $D=1$, $K=1$, $J=1$ and $V=1$. The numerical  data are presented up to the distance $x$, where the finite size of the system does not affect the results. }
\label{convection}
\end{figure}

\begin{figure}
\begin{center}
\includegraphics[width=8cm, height=5.0cm]{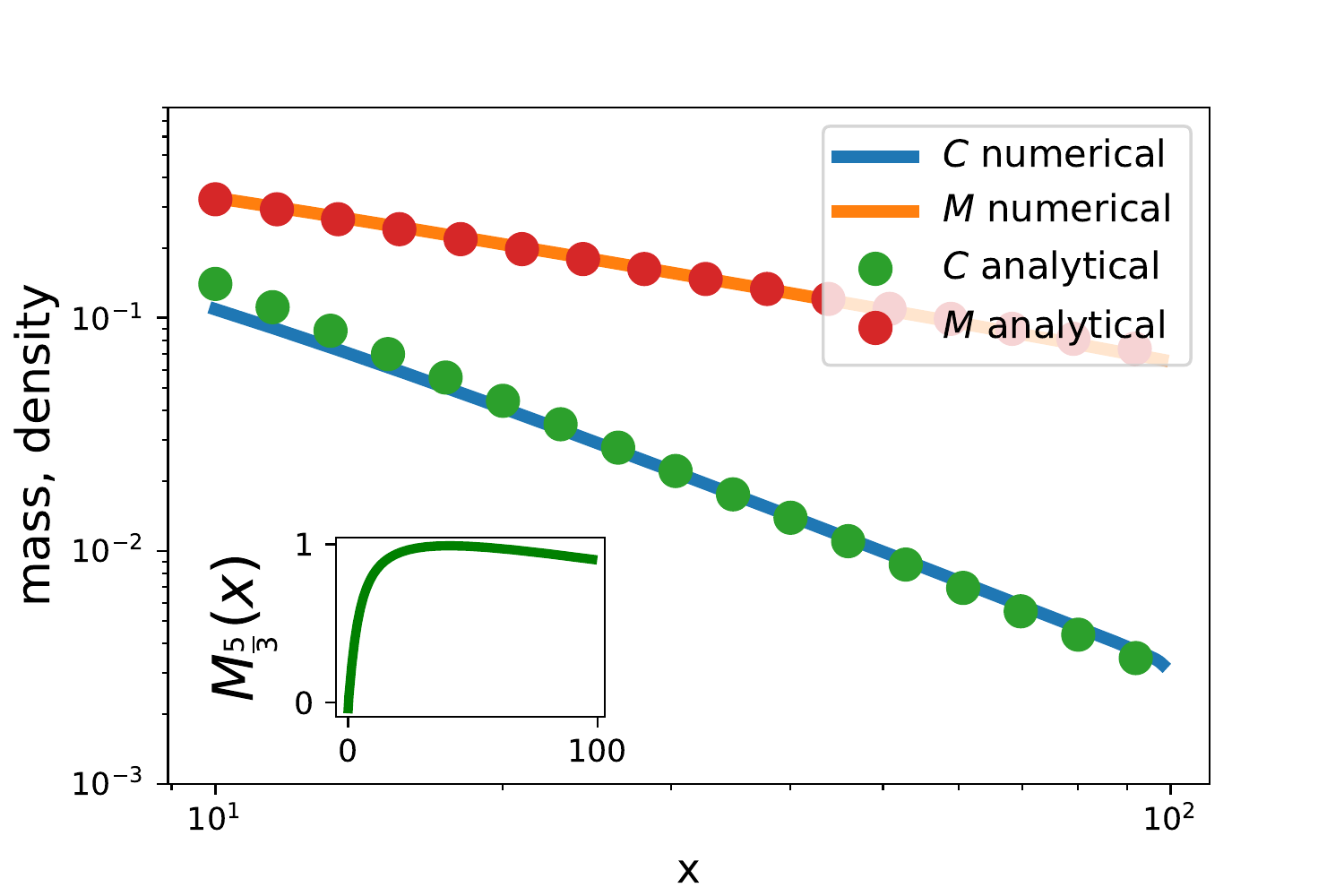}
\end{center}
\caption{Steady-state solutions ($t = 200$) for  the  diffusion-advection-coagulation system \eqref{loc:input} with mass  dependent coefficients: The total mass $m(x)$ and total concentration $c(x)$ as the functions of the distance $x$ from the origin. Lines -- numerical results, dots -- the scaling theory, Eqs. \eqref{cx}, \eqref{mx} with $C_0=6.5$ and $C_1=1.5$ as the fitting constants. The inset plot indicates the dependence of $M_{5/3}$ on $x$.}
\label{size_dependent_5}
\end{figure}

\begin{figure}
\includegraphics[width=8cm, height=7.0cm]{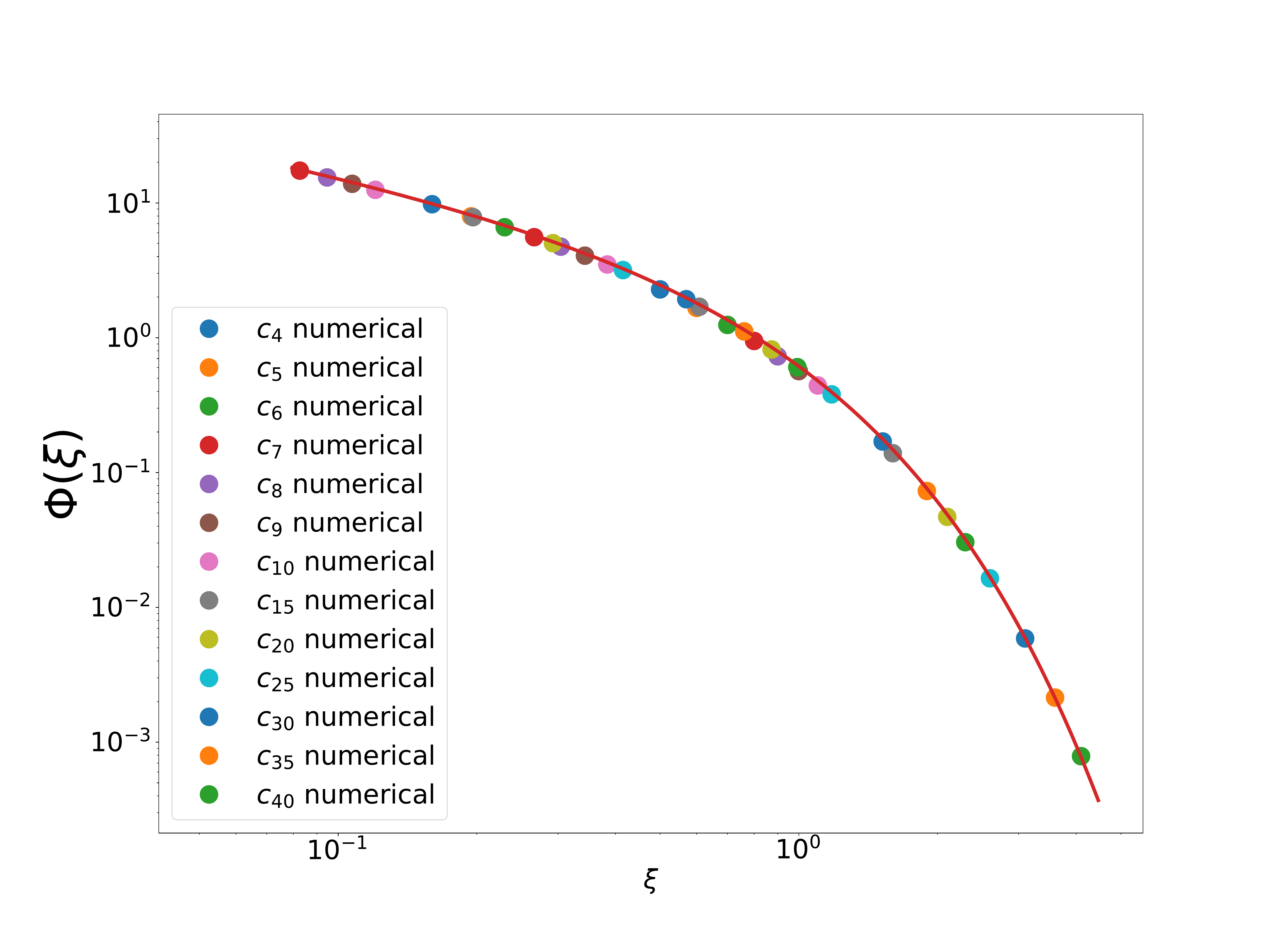}
\caption{Steady-state solutions ($t = 200$) for  the  diffusion-advection-coagulation system \eqref{loc:input} with mass  dependent coefficients. $x^{8/3}c_k(x)$ is plotted as a function of $\xi=k/x$ for different values of $k$. The collapse of the dependencies on the same curve, corresponding to the scaling function $\Phi(x)$ proves the validity of the scaling theory, Eq. \eqref{scaling-sediment}.}
\label{size_dependent_3}
\end{figure}

\begin{figure}
\begin{center}
\includegraphics[width=8cm, height=7.0cm]{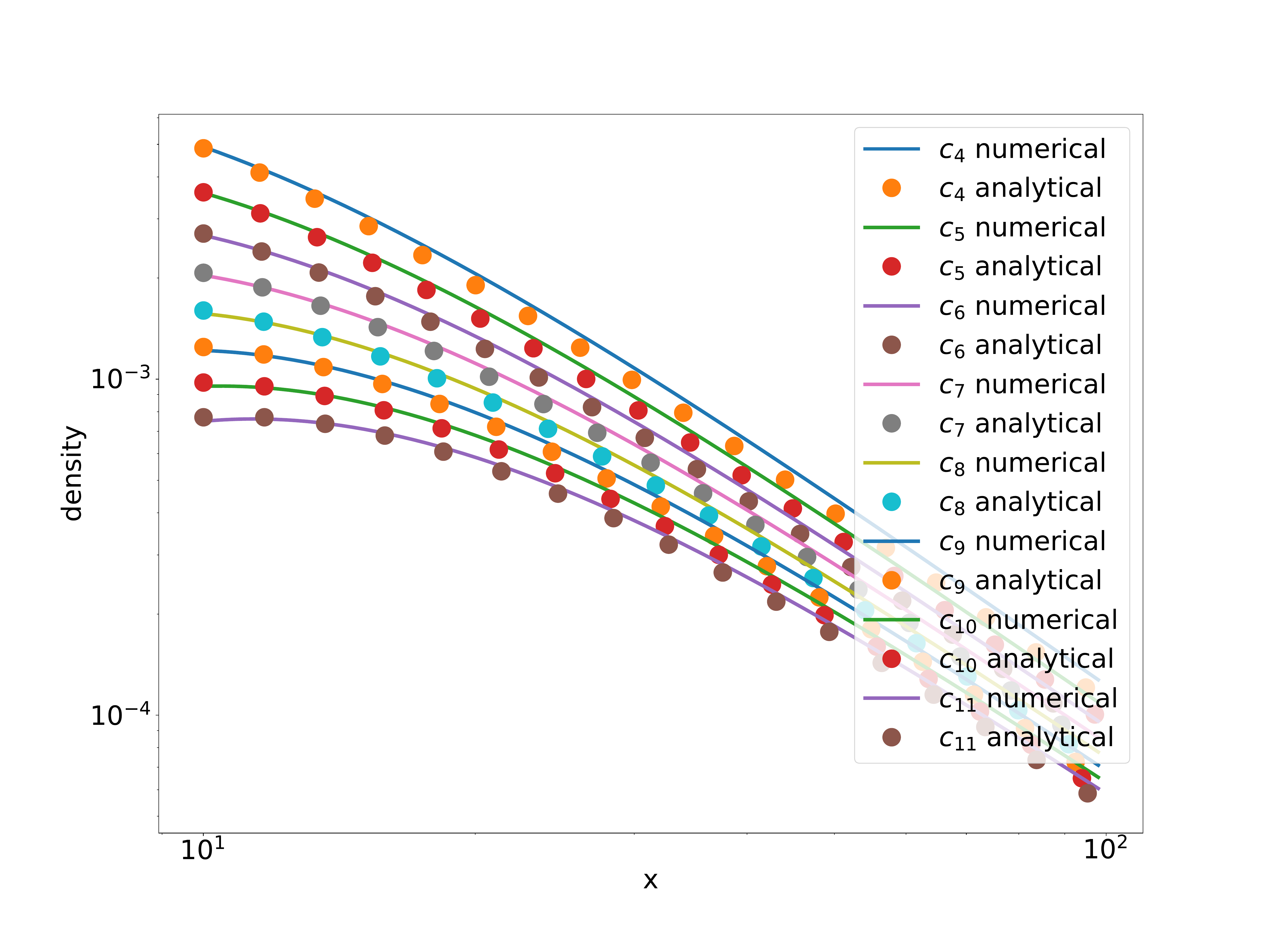}
\end{center}
\caption{Steady-state solutions ($t = 200$) for  the  diffusion-advection-coagulation system \eqref{loc:input} with mass  dependent coefficients. Lines -- numerical results, dots -- the scaling solution, Eqs. \eqref{scaling-sediment}, \eqref{Philast}. The data is fitted for large $\xi = k/x  \gg 1$, where $\Phi =a \xi^{-2/3} e^{-B \xi}$, with the fitting constants $a=3.5$ and $B=1.8$. }
\label{size_dependent_1}
\end{figure}

\begin{figure}
\begin{center}
\includegraphics[width=8cm, height=7.0cm]{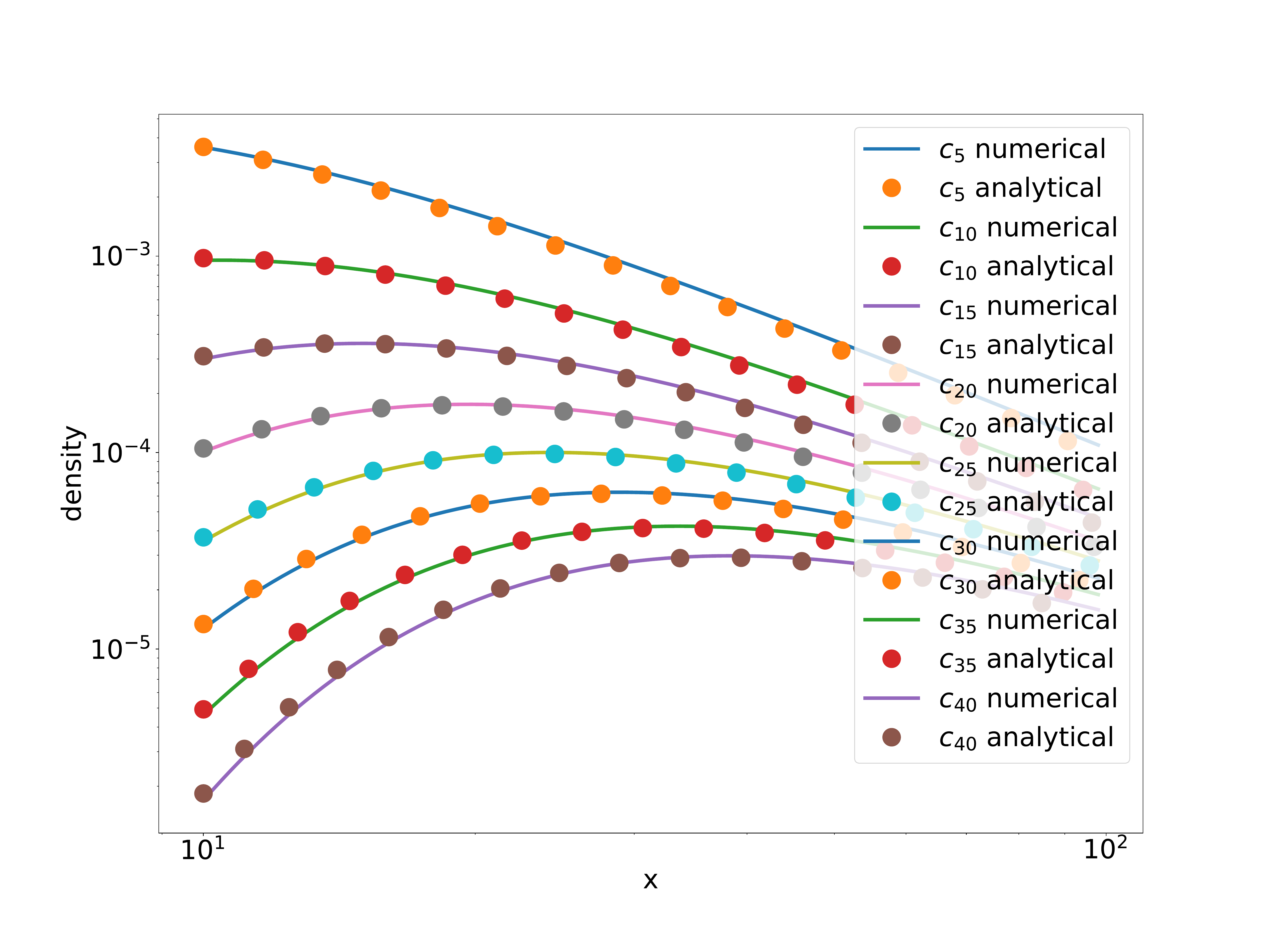}
\end{center}
\caption{The same as in Fig. \ref{size_dependent_1}, but for larger clusters.}
\label{size_dependent_4}
\end{figure}


To check the predictions of the analytical theory we solved the set of Smoluchowki equations \eqref{loc:input} numerically for different coagulation kernels. The coagulation operator was calculated explicitly as it is written in the  right-hand side of Eq. \eqref{loc:input}. The diffusion operator has been implemented in the standard way \cite{Vabishchevich}. The monomer source was modeled by the second order boundary conditions according to Eq.~\eqref{BC}: 
$c_1^{\prime}(+0)= -J/2$ and $c_k^{\prime}(+0)= 0$ for $k \geq 2$. For the advection operator an implicit scheme has been exploited. Additionally, we applied the following modification to the advection:
$$
\Delta_x = \frac{\partial}{\partial x}\left(\frac{q(x)}{p(x)} \frac{\partial [p(x) c_k(x)]}{\partial x} \right)
$$
The above transformation was needed to stabilize the numerical error that occurs for high P\'{e}clet numbers (proportional to the ratio of advection velocity and diffusion); the numerical operator cannot control the error  in this case.  With the new structure the numerical operator acquires the terms suppressing the destabilizing effect of a too intensive flow; this done  by taking smaller discretization steps \cite{Vabishchevich}. In our case 
\begin{equation}
\notag
\Delta_x = \partial_x\!\left(D_k\, e^{\frac{V_k}{D_k} x}\,\, \partial_x\!\left[e^{-\frac{V_k}{D_k} x}
c_k(x)\right]\right)
\end{equation}
In a lack of any computational tricks, the expected computation complexity of  each time-step may be estimated as $O(M N^2)$ -- this number of  operations is needed to simulate $N$ different clusters sizes at $M$  spatial grid points. Indeed, for each grid point $O(N^2)$ operations are requested for a straightforward evaluation of the aggregation sums. Such a complexity means that   computations for the grid $M \times N$ are already very challenging for $M = 1000$ and $N = 256$.  

To overcome the computational difficulty of the solution of large systems of Smoluchowski equations, an efficient method of
low-rank approximation for coagulation kernel has been recently developed \cite{matveev2015, matveev2018anderson,
skorych2019investigation, osinsky2020low}. It requires only  $O(N \log N) $ operations for each time step for a spatially
uniform system and  $O(M N \log N) $ for a nonuniform one.  \cite{zagidullin_2017}. That is, the low-rank technique
guarantees a dramatic increase in the computational speed for large systems of equations. The method is applicable when
the kernel $K_{i,j}$ allows the low-rank decomposition. The class of such kernels is rather broad and includes
a constant kernel, kernels of the type $i^aj^b$, their combinations, and many others. Here we exploit the low-rank
decomposition technique.

The main idea of this approach comes from the reduction of complex summation operations with coagulation kernel to basic
linear convolutions and matrix by vector multiplications. The detailed cross-validation of the accuracy and efficiency of
the new method in comparison with coarse-graining and Monte Carlo approaches has been reported in our previous works
\cite{matveev2018anderson, osinsky2020low, kalinov2021direct}.  Moreover, in the case of Eqs.~\eqref{loc:input} and the
respective simple generalizations, such fast algorithms can be utilized in parallel; a good parallel scaling may be
achieved at modern computing clusters \cite{matveev2018parallel,zagidullin2019supercomputer} supported by thousands of CPU-cores and GPU accelerators.

The results of the numerical simulations are illustrated in Figs.~\ref{diffusion}--\ref{size_dependent_4}, where these are
compared with the predictions of the analytical theory. In Figs.~\ref{diffusion}--\ref{convection} the steady-state
densities of clusters of different sizes $c_k$ and the total density $c$ for the case of constant kinetic coefficients are
shown as the functions of the distance $x$ from the origin (where the monomer source is located). For the case of
vanishing advection, $V=0$, Fig. \ref{diffusion}, and for the case when the advection presents, Fig. \ref{convection}, an
excellent agreement between the theory and simulations is observed.

Next, we perform simulations for the case of mass-dependent coefficients, as given by Eqs.~\eqref{Vk} and \eqref{Dk} with
$V_1=1$ and $D_1=1$. Unfortunately, the ballistic kernel,  \eqref{Kball} or \eqref{Kxinu} does not allow the low-rank
approximation  (due to the presence of the factor with the modulus). Still, in the scaling regime, the solution is
determined mainly by the homogeneity exponents of the kernel $\lambda$ and $\nu$. Moreover, the results of the above
scaling theory do not rely on the particular form of the coagulation kernel, but only on the homogeneity exponent. This
motivates us to perform the numerical simulations for the kernel $K_{i,j} =(ij)^{2/3}$, which has the same homogeneity
exponents $\lambda$ and $\nu$ as the ballistic kernel. Similarly, this yields the same exponents $\lambda =\nu=2/3$ for the
kernels $K_{i,j}^{(1)}$ and $K_{i,j}^{(2)}$ in Eq.~\eqref{KK}. Such a kernel allows the low-rank decomposition and hence
efficient computations.

The results for the mass-dependent coefficients are shown in Figs. \ref{size_dependent_5} -- \ref{size_dependent_4}. In Fig. \ref{size_dependent_5}, the numerical dependence  of the steady-state total density and mass are  compared with the analytical one. The figure clearly indicates the validity of the scaling theory. Further confirmation of the scaling theory is given in Fig. \ref{size_dependent_3}. Here the dependencies $x^{8/3}c_k(x)$ on $\xi =k/x$, collapse on the same scaling function $\Phi (\xi)$ for all $k$, as predicted by the scaling theory, Eq. \eqref{scaling-sediment}.  Note the lack of any fitting parameters in this case. Next, one can also check the theoretical prediction for the form of the scaling function $\Phi (\xi)$, Eq. \eqref{Philast}. This is illustrated in Figs. \ref{size_dependent_1} and \ref{size_dependent_4}. Here the numerical dependence for the densities $c_k(x)$ is compared with the theoretical, Eqs. \eqref{scaling-sediment}, \eqref{Philast} for  large $\xi = k/x  \gg 1$. Again, a very good agreement between the simulation results and the theory is observed. 

Finally, we have checked the prediction of the local conservation of the moment $M_{5/3}$, which is expected to hold when $V^{-1} \ll x \ll (Jt)^3$. We have observed that in the  interval $15<x <100$, the moment depends on $x$ only slightly, varying in the diapason of about 10\%, see the inset in Fig. \ref{size_dependent_5}. These observations support the scaling theory relying on the conservation of $M_{5/3}$. 

\section{Conclusion}
\label{sec:conclusion}

We have performed analytical and numerical analysis of aggregating systems with diffusion of particles, advection and a source of monomers. Mathematically, the system is described by partial differential equations of Smoluchowski type, where the terms responsible for advection and diffusion of aggregates are added. In this work we consider quasi-one dimensional systems, with the space inhomogeneity only in one direction, say $x$, and with the source of monomers localized at $x=0$. The presence of the monomer source entails the appearance of quasi-stationary solutions. Such solutions do not depend on time within some length interval, which expands unlimitedly  in the course of time. Firstly, we have considered the simplest model of constant kinetic coefficients, for both transport  and reaction rate coefficients, governing the aggregation. For this model we have derived exact analytical solution for the density  of agglomerates of different size as the function of space coordinate. Secondly, we have considered a more realistic model with mass-dependent  coefficients in the context of sedimentation process. We have shown that for the stationary case, the diffusion-advection-coagulation equations may be reduced to the time-dependent Smoluchowski-like equations for a spatially homogeneous system. We also demonstrate, that in contrast to the conventional Smoluchowski equations, where the total mass of clusters is conserved, a quasi-conservation law holds in our systems for the moment of the order $5/3$. Namely, the $5/3$-order moment of the cluster-size distribution function is spatially independent for a space interval; the length of this interval  infinitely grows with time. Based on the quasi-conservation law we develop a scaling theory for the densities of agglomerates and their mass. We also derive asymptotic relations for the scaling function describing the cluster densities. 

To check the predictions of the analytical theory we have performed comprehensive numerical simulations and find numerically the spatially dependent agglomerate densities. As the problem is computationally challenging, we apply the effective numerical scheme of low-rank approximation for the reaction kernel of the Smoluchowski equations and adopt this method for the case when   diffusion and advection present. This allows  investigate large systems of PDE equations, that is, the systems with a large space grid and a large number of aggregation equations in each  spatial grid point. 

We have compared the numerical solutions with the predictions of the theory. We find an excellent agreement between the simulation data and exact solutions and confirm the validity of the scaling theory. 

\section*{Acknowlegement}  
This work was partly funded by the Russian Science Foundation, grant No. 21-11-00363. Simulations were performed using Skoltech computational facilities, particularly Zhores \cite{zacharov2019zhores}.

\bibliography{main}

\end{document}